\documentclass{appolb}
\usepackage{graphicx}

\begin{document}
\title{Dynamical kurtosis of net and total proton distributions in STAR at RHIC
}
\author{Zhiming Li
\\
(for the STAR Collaboration)
\address{Key Laboratory of Quark and Lepton Physics (MOE) and Institute of Particle
Physics, Central China Normal University, Wuhan 430079, China}
}
\maketitle
\begin{abstract}
We report the energy and centrality dependence of dynamical kurtosis for Au + Au collisions at $\sqrt{s_{NN}}$ = 7.7, 11.5, 19.6, 27, 39, 62.4 and 200 GeV at RHIC. The dynamical kurtosis of net-proton is compared to that of total-proton. The results are also compared with AMPT model calculations.
\end{abstract}

\section{Introduction}
Mapping the QCD phase diagram as a function of temperature ($T$) and baryon
chemical potential ($\mu_{B}$) is one of the main goals of current heavy ion
experiments. The phase transition at vanishing $\mu_{B}$ is believed to be a crossover~\cite{Crossover}.
Model calculations suggest that the phase transition at large baryon chemical potential is of first order~\cite{firstorder}. There should exist a point where the first-order phase
transition ends, which we call the QCD critical point (QCP)~\cite{QCP1, QCP2}.
For the infinite thermodynamical system
near the QCP, the correlation length ($\xi$) goes to infinity and
the long range correlations become dominant. Whereas, the
correlation length can not be fully developed in current relativistic
heavy-ion experiments due to the finite system size and finite
evolution time. So the higher cumulant ($\sigma = \sqrt{\langle (N-\langle N\rangle)^{2}\rangle},
S = \frac{\langle(N-\langle N\rangle)^{3}\rangle}{\sigma^{3}},
\kappa = \frac{\langle(N-\langle N\rangle)^{4}\rangle}{\sigma^{4}} - 3$) that are more
sensitive to correlation length are suggested~\cite{effective,cumulant}. Here, $N$ can be the number of baryons in an event. Theoretical calculations of the location and property of the QCP meet numerical problems. In experimental aspect, the RHIC beam energy scan program~\cite{bes1, bes2} has been motivated to search for the signature of critical point by scan the collision energies in a wide range. It is shown that
the fluctuations of proton number can reflect the singularity of the baryon number susceptibility as expected at the QCP~\cite{netproton}. So the higher cumulant of proton number is used in current heavy ion experiments~\cite{STARorder}.

In experiment, the numbers of measured protons and antiprotons at RHIC energies are still small. In calculation of the higher cumulants, the statistical fluctuations are not negligible and should be subtracted from directly measured cumulants~\cite{dynamicalJPG}
\begin{equation}
\kappa_{\rm dyn}=\kappa-\kappa_{\rm stat}.
\end{equation}

If we have two independent Poisson distributions for proton and antiproton, respectively, then the net-proton number will follow a Skellam distribution, and the total-proton number will obey a new Poisson. The statistical contributions of kurtosis  for both  net-proton and total-proton are equal to $1/(\langle N_{p}\rangle+\langle N_{\bar{p}}\rangle)$. They are determined only by the means of proton and antiproton in the selected event sample.

By describing fluctuations of the order parameter field $\sigma$ near the critical point, the calculations of the Sigma model predict that the dynamical kurtosis is universally negative when the critical point is approached from the crossover side of the phase  separation line~\cite{sigma}. The negative dynamical kurtosis should be firstly observed in more peripheral collisions and sign change at lower incident energies.

\section{Data sample}
In this work, we use the data of Au+Au collisions at $\sqrt{s_{NN}} =$ 7.7, 11.5, 19.6, 27, 39, 62.4 and 200 GeV collected by the STAR experiment at RHIC.

Our analysis is restricted to particles measured within the STAR
Time Projection Chamber (TPC) detector. The protons and
antiprotons are identified with the ionization energy loss ({\it dE/dx})
measured by the TPC within mid-rapidity of $|y|<0.5$ and
$0.4<p_{T}<0.8$ GeV/$c$. Centrality is defined by the number of uncorrected charged particles within pseudorapidity $|\eta|<0.5$, but excluding protons
and antiprotons to suppress the auto-correlation effects. For each centrality, the mean value of the number of participants is calculated from the Glauber model simulations.

The centrality bin width correction is used to eliminate the volume fluctuation effect~\cite{cbwe}. The statistical error is calculated from the Delta theorem method~\cite{deltamethod}. The systematic uncertainties are estimated by varying the track quality cuts and particle identification for protons and antiprotons.

\section{Results}
\begin{figure}[htb]
\centerline{%
\includegraphics[height=2.6in]{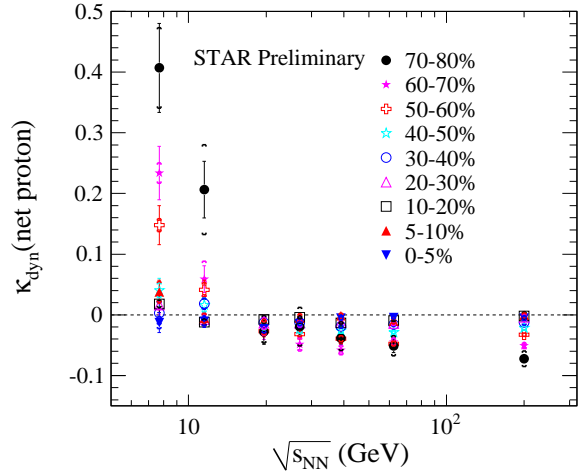}}
\caption{(Color online) Dynamical kurtosis of net-proton as a function of collision energy in nine centrality bins. The brackets represent the systematic uncertainties.}
\label{Fig:F2H}
\end{figure}

The energy dependence of the dynamical kurtosis of net-proton in nine different centrality bins is plotted in Fig.~1. The solid circles represent the most peripheral (70-80$\%$) collisions. The solid reversed triangles represent the most central (0-5$\%$) collisions. We observe that below 19.6 GeV the dynamical kurtosis in peripheral collisions are all positive within errors. Its value increases towards more peripheral collisions. But above 19.6 GeV, the dynamical kurtosis turns to be negative in peripheral collisions.

\begin{figure}[htb]
\centerline{%
\includegraphics[height=2.8in]{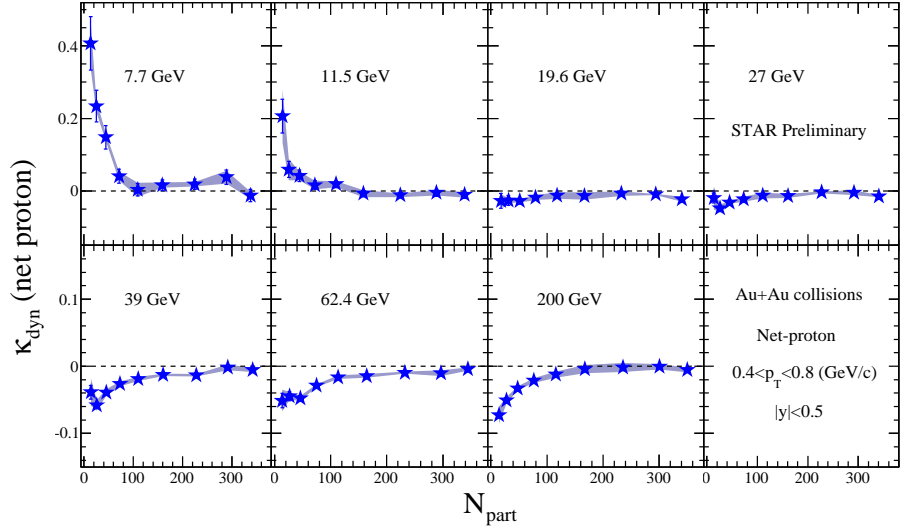}}
\caption{(Color online) Centrality dependence of dynamical kurtosis of net-proton for Au+Au collisions from $\sqrt{s_{NN}} =$ 7.7 to 200 GeV. The shadowed areas represent the systematic uncertainties.}
\label{Fig:F2H}
\end{figure}

Figure 2 shows the centrality dependence of the dynamical kurtosis of net-proton in different RHIC energies. In agreement with Fig.~1, we find that in peripheral collisions the sign of the dynamical kurtosis of net-proton changes with incident energy and collision centrality.

\begin{figure}[htb]
\centerline{%
\includegraphics[height=2.6in]{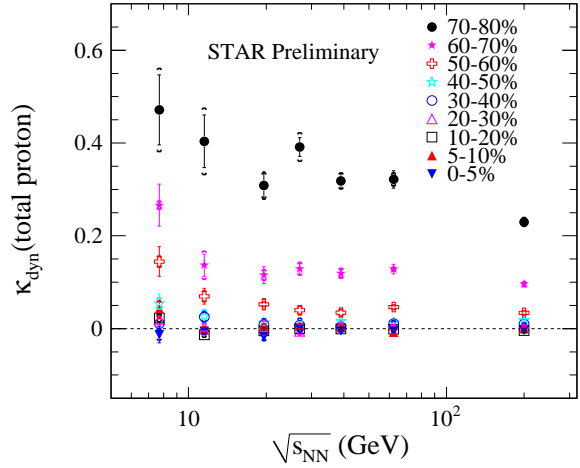}}
\caption{(Color online) Dynamical kurtosis of total-proton as a function of collision energy in nine centrality bins. The brackets represent the systematic uncertainties.}
\label{Fig:F2H}
\end{figure}

\begin{figure}[htb]
\centerline{%
\includegraphics[height=2.8in]{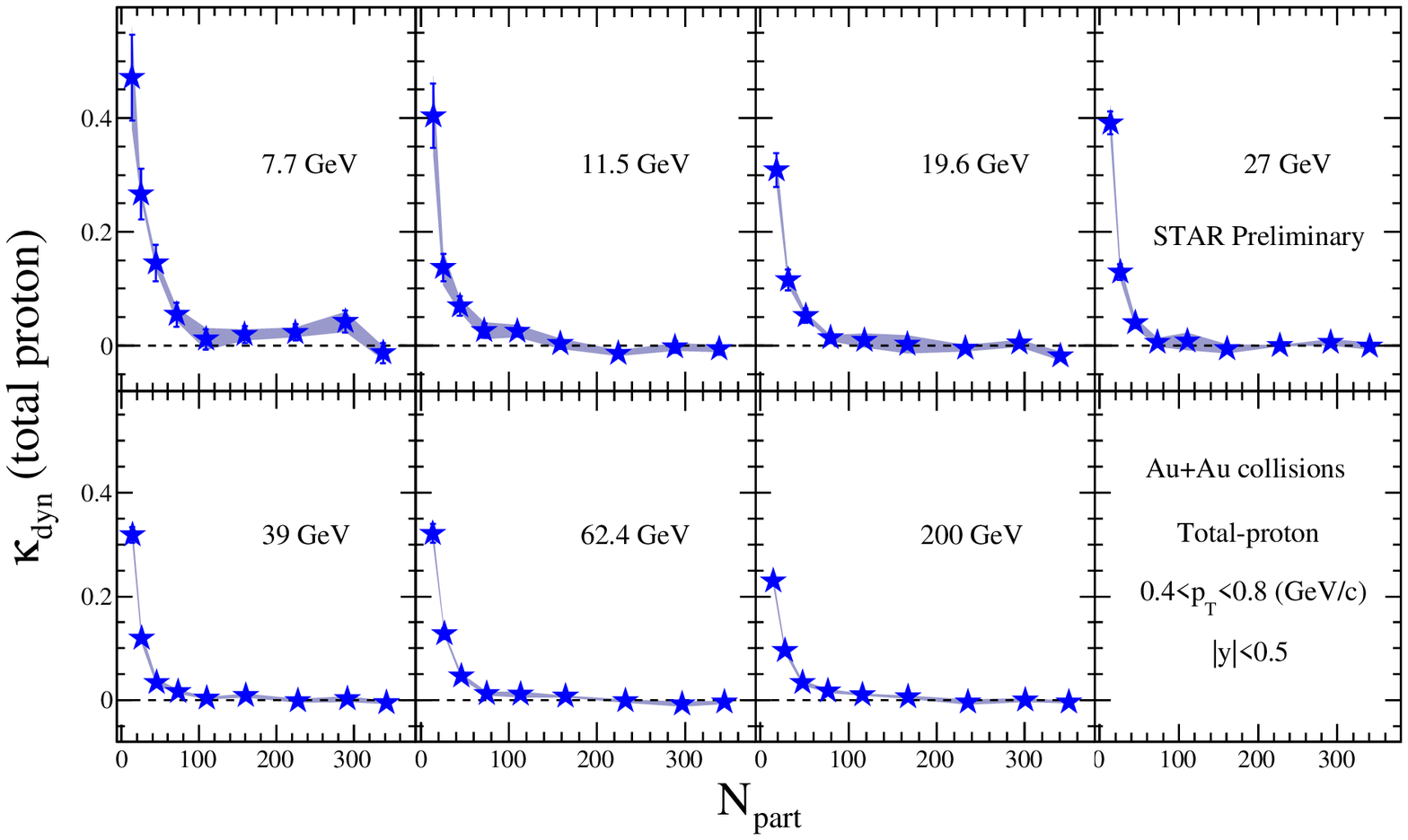}}
\caption{(Color online) Centrality dependence of dynamical kurtosis of total-proton for Au+Au collisions from $\sqrt{s_{NN}} =$ 7.7 to 200 GeV. The shadowed areas represent the systematic uncertainties.}
\label{Fig:F2H}
\end{figure}

It is known that net-baryon number is a conserved quantity, but total-baryon number is not. To see whether the observed sign behavior is particular to the conserved charge, we measure the dynamical kurtosis of total-proton as a function of collision energy and centrality in Fig.~3 and Fig.~4, respectively. The dynamical kurtosis of total-proton in peripheral collisions is found to be always positive at all measured energies. In central collisions, its value is around zero at all incident energies. Therefore, in contrary to the dynamical kurtosis of net-proton in peripheral collisions, we do not observe a sign change behavior for that of total-proton.

\begin{figure}[htb]
\centerline{%
\includegraphics[height=2.1in]{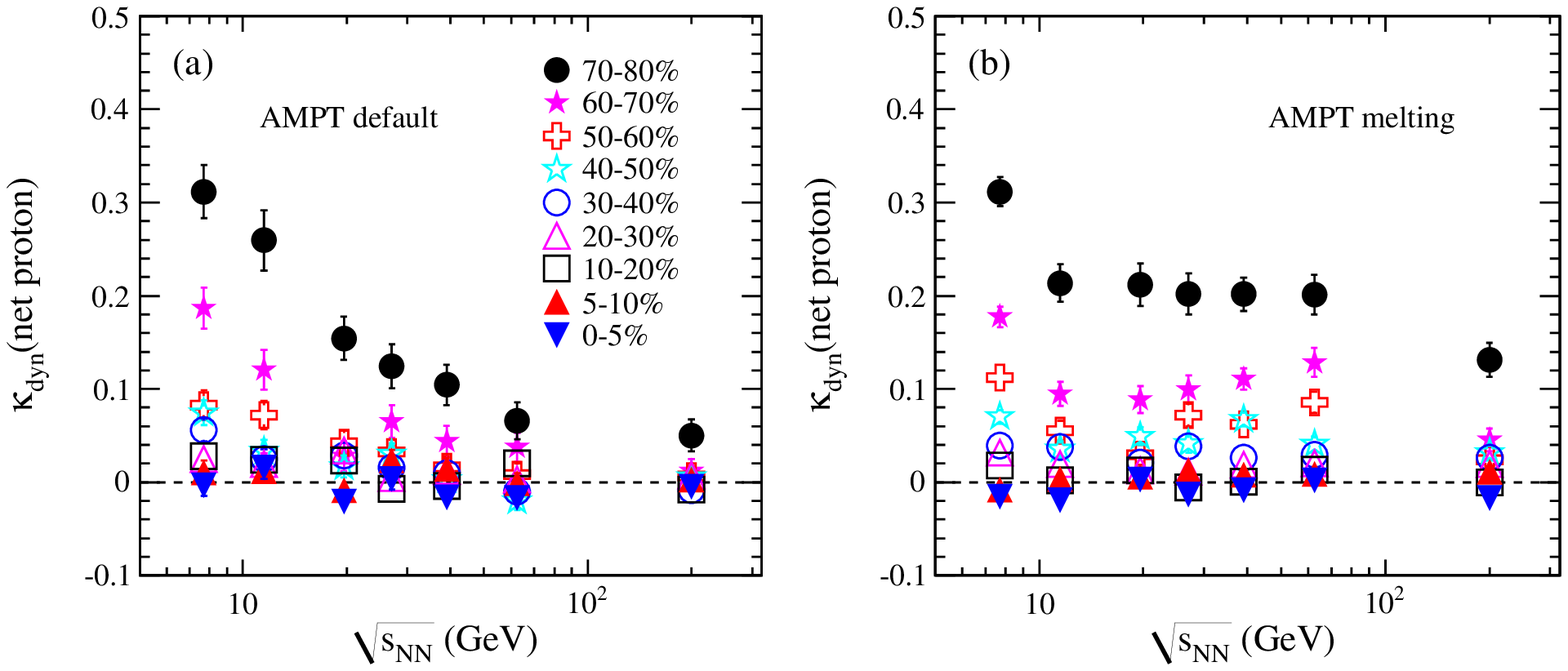}}
\caption{(Color online) Dynamical kurtosis of net-proton as a function of collision energy from AMPT models with (a) default and (b) string melting versions.}
\label{Fig:F2H}
\end{figure}

In order to see if conventional non-critical effects can lead to the sign behavior observed in Fig.~1, we calculate the dynamical kurtosis of net-proton based on the transport model of AMPT~\cite{ampt}. To make our calculations convenient for comparison with the STAR experiment, we choose the same kinematic cuts as used in data for protons and antiprotons. The energy dependence of kurtosis from AMPT models with default version and string melting are shown in Fig.~5. We observe the dynamical kurtosis is positive in non-central collisions at all incident energies. No sign change is observed for two settings of the AMPT models.

\section{Summary}
The dynamical kurtosis has been measured in Au+Au collisions at $\sqrt{s_{NN}}$ = 7.7 to 200 GeV at STAR experiment. In peripheral collisions, the sign of dynamical kurtosis of net-proton is found to change from negative to positive when incident energy decreases. In the contrary, the sign of that of total-proton in peripheral collisions keeps positive at all measured energies. From AMPT model calculations where no critical behavior is included, the dynamic kurtosis for net-proton is found to be positive in non-central collisions for all energies.

\section{Acknowledgement}
This work was supported in part by the National Natural Science Foundation of China under grant No.
11005046 and 10835005. CCNU-QLPL Innovation Fund (QLPL2011P01).


\begin{thebibliography}{99}
\bibitem{Crossover}Y. Aoki {\it et al.}, Nature {\bf 443},
675 (2006); M. Cheng {\it et al.}, Phys. Rev {\bf D 74}, 054507
(2006).

\bibitem{firstorder}E.S. Bowman and J.I. Kapusta, Phys. Rev. {\bf C
79}, 015202 (2009); S. Ejiri, Phys. Rev. {\bf D 78}, 074507 (2008).

\bibitem{QCP1}M.A. Stephanov, Prog. Theor. Phys. Suppl. {\bf 153},
139 (2004); Z. Fodor and S.D. Katz, JHEP {\bf 0404}, 50 (2004).

\bibitem{QCP2}R.V. Gavai and S. Gupta, Phys. Rev. {\bf D 78},
114503 (2008); Phys. Rev. {\bf D 71}, 114014 (2005).

\bibitem{effective}M. Asakawa, S. Ejir, and M. Kitazawa, Phys. Rev.
Lett. {\bf 103}, 262301 (2009).

\bibitem{cumulant} V. Koch, arXiv:0810.2520.

\bibitem{bes1} M.M. Aggarwal {\it et al.} (STAR Collaboration), Phys. Rev. Lett. {\bf 105} 022302 (2011).

\bibitem{bes2} M.M. Aggarwal {\it et al.} (STAR Collaboration), arXiv:1007.2613.

\bibitem{netproton} Y. Hatta and A. Stephanove, Phys.
Rev. Lett. {\bf 91}, 102003 (2003); M. Kitazawa and M. Asakawa, arXiv:1107.2755.

\bibitem{STARorder} M.M. Aggarwal {\it et al.} (STAR Coll.), Phys. Rev. Lett. {\bf 105},
022302 (2010).

\bibitem{dynamicalJPG}L. Chen {\it et al.}, J Phys. G: Nucl. Part. Phys. {\bf 38}, 115004 (2011).

\bibitem{sigma}M.A. Stephanov, Phys. Rev. Lett. {\bf 107},
052301 (2011); Phys. Rev. Lett. {\bf 102}, 032301 (2009).

\bibitem{cbwe} X. Luo (for the STAR Collaboration), J. Phys.:Conf. Ser. {\bf 316}, 012003 (2011).

\bibitem{deltamethod} X. Luo (for the STAR Collaboration), J. Phys G: Nucl. Part. Phys. {\bf 39}, 025008 (2012).

\bibitem{ampt} Z.W. Lin, C.M. Ko, B.A. Li {\it et al.},
Phys. Rev. {\bf C 72}, 064901 (2005).

\end{thebibliography}
\end{document}